\documentclass[a4paper, 11pt]{article}
\usepackage[utf8]{inputenc} 

\usepackage{amssymb,amsfonts,amsmath,mathtext,cite,enumerate,float,amsthm}
\usepackage{ytableau}
\usepackage[vcentermath]{youngtab}
\usepackage{tikz}\usetikzlibrary{tikzmark}
\usetikzlibrary{fadings} 
\usetikzlibrary{positioning}
\usetikzlibrary{backgrounds} 
\usetikzlibrary[backgrounds]
\usetikzlibrary{decorations.pathreplacing}
\usetikzlibrary{arrows}
\graphicspath{ {C:/Users/yourf/OneDrive/Documents/Наука/ТеХ/} }
\usepackage{tikz-cd}
\usepackage{genyoungtabtikz}
\usepackage{hyperref}
\usepackage{caption}
\hoffset= - 1.0in

\theoremstyle{definition}

\theoremstyle{plain}

\theoremstyle{remark}
\newtheorem*{Remark}{Remark}

\newcommand{\sln}{\mathfrak{sl}}
\newcommand{\qdim}{\operatorname{qdim}}
\newcommand{\eqnb}{\begin{equation}}
\newcommand{\eqn}{\end{equation}}

\newcommand{\h}{\hbar}

\newcommand{\K}{\mathcal{K}}
\newcommand{\T}{\textbf{T}}

\newcommand{\A}{\mathcal{A}}
\newcommand{\Hn}{\mathcal{H}}
\newcommand{\R}{\mathcal{R}}

\definecolor{airforceblue}{rgb}{0.36, 0.54, 0.66}	\definecolor{beige}{rgb}{0.96, 0.96, 0.86}
\definecolor{bittersweet}{rgb}{1.0, 0.44, 0.37}
\definecolor{melon}{rgb}{0.99, 0.74, 0.71}
\definecolor{mustard}{rgb}{1.0, 0.86, 0.35}
\definecolor{lava}{rgb}{0.81, 0.06, 0.13}
\definecolor{magnolia}{rgb}{0.97, 0.96, 1.0}
\definecolor{lavendermist}{rgb}{0.9, 0.9, 0.98}
\definecolor{lavendergray}{rgb}{0.77, 0.76, 0.82}
\definecolor{palepink}{rgb}{0.98, 0.85, 0.87}
\definecolor{palesilver}{rgb}{0.79, 0.75, 0.73}
\definecolor{cadetgrey}{rgb}{0.57, 0.64, 0.69}
\definecolor{anti-flashwhite}{rgb}{0.95, 0.95, 0.96}
\colorlet{Light0anti-flashwhite}{anti-flashwhite!70!white}
\colorlet{Lightanti-flashwhite}{anti-flashwhite!50!white}
\colorlet{Light2anti-flashwhite}{anti-flashwhite!30!white}

\begin{document}

\title{\vspace{0.1cm}{\Large {\bf 
A novel symmetry of colored HOMFLY polynomials coming from $\mathfrak{sl}(N|M)$ superalgebras
}
\date{}
\author{
{\bf V. Mishnyakov $^{a,c,d}$}\thanks{mishnyakovvv@gmail.com},
{\bf A. Sleptsov$^{a,b,d}$}\thanks{sleptsov@itep.ru}, 
{\bf N. Tselousov$^{a,d}$}\thanks{tselousov.ns@phystech.edu}}
}}
\maketitle
\vspace{-5.9cm}

\begin{center}
\hfill \\
\end{center}

\vspace{4.2cm}

\begin{center}

$^a$ {\small {\it Institute for Theoretical and Experimental Physics, Moscow 117218, Russia}}\\
$^b$ {\small {\it Institute for Information Transmission Problems, Moscow 127994, Russia}}\\
$^c$ {\small {\it Moscow State University, Physical Department, Vorobjevy Gory, Moscow, 119899, Russia }} \\
$^d$ {\small {\it Moscow Institute of Physics and Technology, Dolgoprudny 141701, Russia }}

\end{center}

\vspace{1cm}

\vspace{1cm}

\begin{abstract}
We present a novel symmetry of the colored HOMFLY polynomial. It relates pairs of polynomials colored by different representations at specific values of $N$ and generalizes the previously known "tug-the-hook" symmetry of the colored Alexander polynomial \cite{Mishnyakov:2020khb}. As we show, the symmetry has a superalgebra origin, which we discuss qualitatively. Our main focus are the constraints that such a property imposes on the general group-theoretical structure, namely the $\mathfrak{sl}(N)$ weight system, arising in the perturbative expansion of the invariant. Finally, we demonstrate its tight relation to the eigenvalue conjecture.
\end{abstract}

\vspace{.5cm}

\section{Introduction}
Quantum knot invariants are among the hot topics in mathematical physics today. This is due to their appearance in various context \cite{Ramadevi:1993np,Ooguri:1999bv,BarNatan:1990ps,Gukov:2004hz,Mironov:2011ym,Garoufalidis:2015ewa,Kashaev:1996kc,Wu}. One of the most well studied knot invariants is the colored HOMFLY polynomial. It generalizes the colored Jones, Alexander and the $\mathfrak{sl}(N)$ Reshetikhin-Turaev invariants. Their common property is that they arise from representations of $\mathfrak{sl}(N|M)$ superalgebras \cite{Gorsky:2013jxa,QS}.
\\\\
Commonly the colored HOMFLY polymomial is introduced as an observable in $SU(N)$ Chern-Simons theory on $S^3$\cite{Witten,CS}
\begin{equation}
S_{CS}=\dfrac{2 \pi i k}{k+N}\int_{S^3}\left( A \wedge dA +\dfrac{2}{3} A\wedge A\wedge A \right)
\end{equation}
The natural observables in this topological theory are Wilson loops along a knot $\K$ embedded into $S^3$  carrying arbitrary representations $R$ of the gauge group:
\begin{equation}
W(\K)=\mathrm{tr}_R \left(\mathrm{Pexp} { \oint_{\K} A }\right)
\end{equation} 
The HOMFLY polynomial is the expectation value of a Wilson loop:
\begin{equation}
H_R^\K(q,a)=\langle W(\K)\rangle
\end{equation}
It appears to be a function of two variables, which are expressed in terms of CS theory parameters:
\begin{align}
q=e^{\h} ,\quad a=e^{N\h}, \quad \h=\dfrac{2\pi i k}{k+N}
\end{align}
The parameter $\h$ plays the role of the loop expansion parameter in the path integral.

This value of the Wilson loop is not in fact a polynomial, but it contains a simple rational factor, which is the value of the Wilson loop of the unknot. Hence to obtain proper polynomials one should normalize the HOMFLY invariant by its value on the unknot:
\begin{align}\label{NormalizedHOMFLY}
\Hn^\K_R(q,a)=\dfrac{H^\K_R(q,a)}{H^\bigcirc_R(q,a)}.
\end{align}
There are two "parameters" in the definition above: the knot itself and the representation. This paper is focused on the representation dependence of the HOMFLY polynomial. \\\\
The representation $R$ is given by the corresponding Young diagram:
\begin{equation}
R=[R_1,R_2, \ldots R_n]
\end{equation}
Note, that a choice of any $N$ restricts the set of \emph{physical} Young diagrams. Where by physical we mean those, for which colored Wilson loops can be defined. These have to have $n\leq N$, i.e less than $N$ rows to specify a representation of $SU(N)$. However, the normalized HOMFLY polynomial has a proper limit for larger diagrams with $n>N$(see sec.\ref{4}). Therefore we will consider arbitrary Young diagrams for all $N$.

Several interesting properties of the  colored HOMFLY polynomial are known. Let us list some of them:
\begin{itemize}
\item Rank-level duality of Chern-Simons theory \cite{Naculich:1990hg,Naculich:1990pa,Mlawer:1990uv} also called mirror symmetry  provides the following relation \cite{Liu:2010zs}:
\begin{equation}
\Hn^\K_R(q,a)=\Hn^K_{R^T}(q^{-1},a)
\end{equation}
where $R^T$ is the representation obtained by transposing the Young diagram:
\begin{equation}
\ytableausetup{boxsize=0.6em,aligntableaux = center}
R=\ydiagram{4,2,1} \longleftrightarrow \ydiagram{3,2,1,1,}=R^T.
\end{equation}
\item One can take specific limits in the variables $q$ and $a$. The t'Hooft planar limit means scaling :
\begin{equation}
\h \rightarrow 0, \ N \rightarrow \infty , \ N \h \text{ - fixed}
\end{equation}
which corresponds to  setting $ q =1 $. 
The emerging so called special polynomials have a remarkable property \cite{Mironov:2013oma,Itoyama:2012fq,DuninBarkowski:2011yx,Zhu:2012tm}
\begin{align}
\sigma_R^\K(a)=\Hn_R^\K(q=1,a) \nonumber \\
\sigma_R^\K(a)=\left(\sigma_{[1]}^\K(a)\right)^{|R|}
\end{align}
\item The dual limit $a=1$ or $N=0$ giving the colored Alexander polynomial exhibits a similar property \cite{Itoyama:2012fq,Zhu:2012tm,Mironov:2016mxn}:
\begin{equation}
\A_R(q)=\A_{[1]}(q^{|R|}) , \text{ when }R=[r,1^L]
\end{equation}
\end{itemize}
We are going to discuss \textbf{a new property of this kind}. Sections 2 and 3 are devoted to presenting the property and giving qualitative justification for its validity. In section 4 the connection to the eigenvalue conjecture is exposed. Mainly we demonstrate how to validate this symmetry under the assumption that the eigenvalue conjecture holds. The case of $N=0$, i.e. $a=1$ which has already been well studied \cite{Mishnyakov:2020khb} will serve as our main example of some calculations. \section{Tug-the-hook symmetry.}
We want put forward the following symmetry of the normalized colored HOMFLY polynomial:
\begin{equation}\label{sym}
\boxed{\Hn_R(q,A=q^N)=\Hn_{\T^N_\epsilon(R)}(q,A=q^N)}
\end{equation}
where $\T^N_\epsilon$ is a transformation of Young diagrams.
Let us give an example:
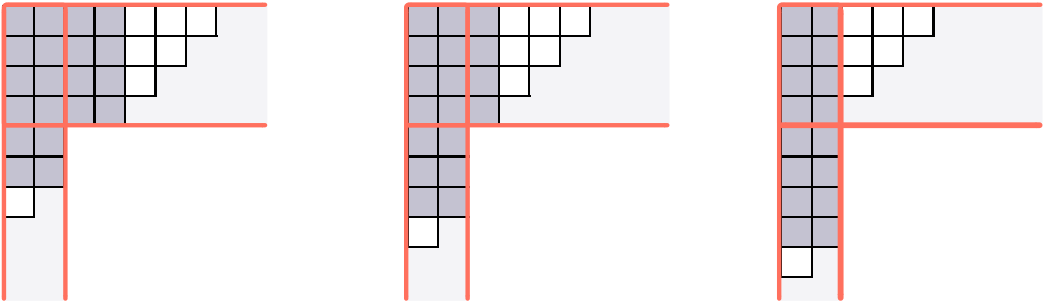
\tikzfading[name=fade right,
left color=transparent!98,
right color=transparent!95]
\tikzfading[name=fade down,
top color=transparent!98,
bottom color=transparent!95]
\begin{center}
\begin{figure}[H]
\centering
\begin{tikzpicture}[>=triangle 45,font=\sffamily,rounded corners=1pt]
    \node  (1) at (0,0) {$\ytableausetup
{boxsize=1em}
\ytableausetup{aligntableaux = top}
\ydiagram[*(white)]{4+3,4+2,4+1,0,0,0,1}*[*(lavendergray)]{4,4,4,4,2,2} \hspace{2.5cm} \ydiagram[*(white)]{3+3,3+2,3+1,0,0,0,0,1}*[*(lavendergray)]{3,3,3,3,2,2,2} \hspace{2.5cm}  \ydiagram[*(white)]{2+3,2+2,2+1,0,0,0,0,0,1}*[*(lavendergray)]{2,2,2,2,2,2,2,2}$};

\draw[line width=0.15em,bittersweet,round cap-round cap]([xshift=0.4em,yshift=-0.45em]1.north west)--([xshift=0.4em,yshift=-10.55em]1.north west);
\draw[line width=0.15em,bittersweet,round cap-round cap]([xshift=2.5em,yshift=-0.45em]1.north west)--([xshift=2.5em,yshift=-10.55em]1.north west);
\draw[line width=0.15em,bittersweet,round cap-round cap]([xshift=0.4em,yshift=-0.35em]1.north west)--([xshift=9.4em,yshift=-0.35em]1.north west);
\draw[line width=0.15em,bittersweet,round cap-round cap]([xshift=0.4em,yshift=-4.5em]1.north west)--([xshift=9.4em,yshift=-4.5em]1.north west);
\draw [line width=0.15em,bittersweet]([xshift=0.4em,yshift=-0.35em]1.north west) rectangle ([xshift=2.5em,yshift=-4.5em]1.north west);
\begin{scope}[on background layer]
\fill [anti-flashwhite,opacity=0.8]([xshift=0.4em,yshift=-0.35em]1.north west) rectangle ([xshift=2.5em,yshift=-10.55em]1.north west);
\fill [anti-flashwhite,opacity=0.8]([xshift=0.4em,yshift=-0.35em]1.north west) rectangle ([xshift=9.4em,yshift=-4.5em]1.north west);
\end{scope}
\draw[line width=0.15em,bittersweet,round cap-round cap]([xshift=0.4em+7.25em+2.5cm,yshift=-0.45em]1.north west)--([xshift=0.4em+7.25em+2.5cm,yshift=-10.55em]1.north west);
\draw[line width=0.15em,bittersweet,round cap-round cap]([xshift=2.5em+7.25em+2.5cm,yshift=-0.45em]1.north west)--([xshift=2.5em+7.25em+2.5cm,yshift=-10.55em]1.north west);
\draw[line width=0.15em,bittersweet,round cap-round cap]([xshift=0.4em+7.25em+2.5cm,yshift=-0.35em]1.north west)--([xshift=9.4em+7.25em+2.5cm,yshift=-0.35em]1.north west);
\draw[line width=0.15em,bittersweet,round cap-round cap]([xshift=0.4em+7.25em+2.5cm,yshift=-4.5em]1.north west)--([xshift=9.4em+7.25em+2.5cm,yshift=-4.5em]1.north west);
\draw [ line width=0.15em,bittersweet]([xshift=7.65em+2.5cm,yshift=-0.35em]1.north west) rectangle ([xshift=7.6em+2.5em-0.35em+2.5cm,yshift=-4.5em]1.north west);
\begin{scope}[on background layer]
\fill [anti-flashwhite,opacity=0.8]([xshift=0.4em+7.25em+2.5cm,yshift=-0.35em]1.north west) rectangle ([xshift=2.5em+7.25em+2.5cm,yshift=-10.55em]1.north west);
\fill [anti-flashwhite,opacity=0.8]([xshift=0.4em+7.25em+2.5cm,yshift=-0.35em]1.north west) rectangle ([xshift=9.4em+7.25em+2.5cm,yshift=-4.5em]1.north west);
\end{scope}
\draw[line width=0.15em,bittersweet,round cap-round cap]([xshift=0.4em+13.5em+5cm,yshift=-0.45em]1.north west)--([xshift=0.4em+13.5em+5cm,yshift=-10.55em]1.north west);
\draw[line width=0.2em,bittersweet,round cap-round cap]([xshift=2.5em+13.5em+5cm,yshift=-0.45em]1.north west)--([xshift=2.5em+13.5em+5cm,yshift=-10.55em]1.north west);
\draw[line width=0.15em,bittersweet,round cap-round cap]([xshift=0.4em+13.5em+5cm,yshift=-0.35em]1.north west)--([xshift=9.4em+13.5em+5cm,yshift=-0.35em]1.north west);
\draw[line width=0.2em,bittersweet,round cap-round cap]([xshift=0.4em+13.5em+5cm,yshift=-4.5em]1.north west)--([xshift=9.4em+13.5em+5cm,yshift=-4.5em]1.north west);
\draw [ line width=0.15em,bittersweet]([xshift=13.9em+5cm,yshift=-0.35em]1.north west) rectangle ([xshift=13.9em+2.5em-0.4em+5cm,yshift=-4.5em]1.north west);
\begin{scope}[on background layer]
\fill [anti-flashwhite,opacity=0.8]([xshift=0.4em+13.5em+5cm,yshift=-0.35em]1.north west) rectangle ([xshift=2.5em+13.5em+5cm,yshift=-10.55em]1.north west);
\fill [anti-flashwhite,opacity=0.8]([xshift=0.4em+13.5em+5cm,yshift=-0.35em]1.north west) rectangle ([xshift=9.4em+13.5em+5cm,yshift=-4.5em]1.north west);
\end{scope}
\end{tikzpicture}
\caption*{An example of applying the symmetry for $N=2$.The transformation is applied to the \textbf{\textcolor{cadetgrey}{grey}} parts of the diagrams, which is shifted inside a \textbf{\textcolor{bittersweet}{$(4|2)$ fat hook}}. For more particular examples, see \eqref{example1}-\eqref{example4}.}
\end{figure}
\end{center}


 The symmetry works as follows:
\begin{itemize}
\item One identifies a \textcolor{bittersweet}{$(N+M
|M)$ \textbf{fat hook}}, i.e a diagram extending with columns of height $N+M$ to the right and rows of length $M$, such that $R$ fits into the fat hook, as demonstrated in the figure above. For a given $N$ such choice is unique.
\item The transformation amounts to pulling the diagram  inside \textcolor{bittersweet}{\textbf{fat hook}}. The HOMFLY polynomial is invariant with respect to such transformations.
\item The name comes from the analogy with the "tug of war" game.
\end{itemize}
Note, that the transformation changes the overall number of boxes in the diagram. To describe the symmetry quantitatively let us use an analogue of Frobenius notation for Young diagrams:\\
\begin{itemize}
\item Parametrize the first $N$ rows with their lengths $R_i$ for $1\leq i \leq N$.
\item  The rest of the diagram is parametrized by shifted Frobenius variables:
\begin{equation}\label{shiftednot0}
\begin{split}
\alpha_i&=R_i - (i-N)+1 \\
\beta_i &=R'_{i -N}-i+1 
\end{split}
, \qquad i>N
\end{equation}
\end{itemize}
As a result we describe Young diagrams with the following data
\begin{equation}\label{shiftednot}
    [R_1, \ldots, R_N](\alpha_{N + 1}, \ldots, \alpha_{N + k} \, | \, \beta_{N + 1}, \ldots, \beta_{N + k}).
\end{equation}
Then $\T^N_\epsilon$ is the following transformation:
\begin{align}
R_i \longrightarrow R_i-\epsilon \nonumber \\
\alpha_i \longrightarrow \alpha_i-\epsilon \\
\beta_i \longrightarrow \beta_i+\epsilon \nonumber 
\end{align}
where $\epsilon$ is an integer, such that the result is still a Young diagram. In terms of the shifted Frobenius variables \eqref{shiftednot0},\eqref{shiftednot} the diagrams in the example in the figure above are parametrized as follows:
$$
\hspace{10mm} R = [7, 6](5, 3 \, | \, 4, 2) \hspace{15mm} 
\T^2_{-1}(R) = [6, 5](4, 2 \, | \, 5, 3) \hspace{10mm} 
\T^2_{-2}(R) = [5, 4](3, 1 \, | \, 6, 4)
$$
One can easily convince oneself that for a given diagram $R$ and a fixed $N$ there is only one family of related diagrams labeled by $\epsilon$. In particular for $N=0$ it reduces to the tug-the-hook symmetry of the Alexander polynomial \cite{Mishnyakov:2020khb}:
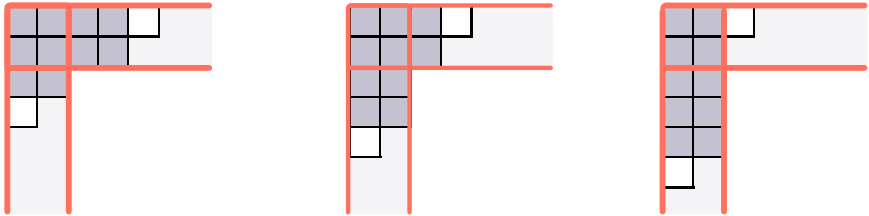
\begin{equation}\label{AlexTTH}
\A^\K_R(q)=\A^\K_{\T_\epsilon(R)}(q)
\end{equation}
\begin{figure}[H]
\centering
\begin{tikzpicture}[>=triangle 45,font=\sffamily,rounded corners=1pt]
    \node (1) at (0,0) {$\ytableausetup
{boxsize=1em}
\ytableausetup{aligntableaux = top}
\ydiagram[*(white)]{4+1,0,0,1}*[*(lavendergray)]{4,4,2} \hspace{2.5cm} \ydiagram[*(white)]{3+1,0,0,0,1}*[*(lavendergray)]{3,3,2,2} \hspace{2.5cm}  \ydiagram[*(white)]{2+1,0,0,0,0,1}*[*(lavendergray)]{2,2,2,2,2}$};

\draw[line width=0.2em,bittersweet,round cap-round cap]([xshift=0.4em,yshift=-0.45em]1.north west)--([xshift=0.4em,yshift=-7.55em]1.north west);
\draw[line width=0.2em,bittersweet,round cap-round cap]([xshift=2.5em,yshift=-0.45em]1.north west)--([xshift=2.5em,yshift=-7.55em]1.north west);
\draw[line width=0.2em,bittersweet,round cap-round cap]([xshift=0.4em,yshift=-0.35em]1.north west)--([xshift=7.4em,yshift=-0.35em]1.north west);
\draw[line width=0.2em,bittersweet,round cap-round cap]([xshift=0.4em,yshift=-2.5em]1.north west)--([xshift=7.4em,yshift=-2.5em]1.north west);
\draw [line width=0.2em,bittersweet]([xshift=0.4em,yshift=-0.35em]1.north west) rectangle ([xshift=2.5em,yshift=-2.5em]1.north west);
\begin{scope}[on background layer]
\fill [anti-flashwhite,opacity=0.8]([xshift=0.4em,yshift=-0.45em]1.north west) rectangle ([xshift=2.5em,yshift=-7.55em]1.north west);
\fill [anti-flashwhite,opacity=0.8]([xshift=0.4em,yshift=-0.45em]1.north west) rectangle ([xshift=7.4em,yshift=-2.5em]1.north west);
\end{scope}
\draw[line width=0.15em,bittersweet,round cap-round cap]([xshift=0.4em+5.15em+2.5cm,yshift=-0.45em]1.north west)--([xshift=0.4em+5.15em+2.5cm,yshift=-7.55em]1.north west);
\draw[line width=0.15em,bittersweet,round cap-round cap]([xshift=2.5em+5.15em+2.5cm,yshift=-0.45em]1.north west)--([xshift=2.5em+5.15em+2.5cm,yshift=-7.55em]1.north west);
\draw[line width=0.15em,bittersweet,round cap-round cap]([xshift=0.4em+5.15em+2.5cm,yshift=-0.35em]1.north west)--([xshift=7.4em+5.15em+2.5cm,yshift=-0.35em]1.north west);
\draw[line width=0.15em,bittersweet,round cap-round cap]([xshift=0.4em+5.15em+2.5cm,yshift=-2.5em]1.north west)--([xshift=7.4em+5.15em+2.5cm,yshift=-2.5em]1.north west);
\draw [ line width=0.15em,bittersweet]([xshift=5.55em+2.5cm,yshift=-0.35em]1.north west) rectangle ([xshift=5.5em+2.5em-0.35em+2.5cm,yshift=-2.5em]1.north west);
\begin{scope}[on background layer]
\fill [anti-flashwhite,opacity=0.8]([xshift=0.4em+5.15em+2.5cm,yshift=-0.45em]1.north west) rectangle ([xshift=2.5em+5.15em+2.5cm,yshift=-7.55em]1.north west);
\fill [anti-flashwhite,opacity=0.8]([xshift=0.4em+5.15em+2.5cm,yshift=-0.45em]1.north west) rectangle ([xshift=7.4em+5.15em+2.5cm,yshift=-2.5em]1.north west);
\end{scope}
\draw[line width=0.2em,bittersweet,round cap-round cap]([xshift=0.4em+9.4em+5cm,yshift=-0.45em]1.north west)--([xshift=0.4em+9.4em+5cm,yshift=-7.55em]1.north west);
\draw[line width=0.2em,bittersweet,round cap-round cap]([xshift=2.5em+9.4em+5cm,yshift=-0.45em]1.north west)--([xshift=2.5em+9.4em+5cm,yshift=-7.55em]1.north west);
\draw[line width=0.2em,bittersweet,round cap-round cap]([xshift=0.4em+9.4em+5cm,yshift=-0.35em]1.north west)--([xshift=7.4em+9.4em+5cm,yshift=-0.35em]1.north west);
\draw[line width=0.2em,bittersweet,round cap-round cap]([xshift=0.4em+9.4em+5cm,yshift=-2.5em]1.north west)--([xshift=7.4em+9.4em+5cm,yshift=-2.5em]1.north west);
\draw [line width=0.15em,bittersweet]([xshift=9.8em+5cm,yshift=-0.35em]1.north west) rectangle ([xshift=9.8em+2.5em-0.4em+5cm,yshift=-2.5em]1.north west);
\begin{scope}[on background layer]
\fill [anti-flashwhite,opacity=0.8]([xshift=0.4em+9.4em+5cm,yshift=-0.45em]1.north west) rectangle ([xshift=2.5em+9.4em+5cm,yshift=-7.55em]1.north west);
\fill [anti-flashwhite,opacity=0.8]([xshift=0.4em+9.4em+5cm,yshift=-0.45em]1.north west) rectangle ([xshift=7.4em+9.4em+5cm,yshift=-2.5em]1.north west);
\end{scope}
\end{tikzpicture}
\caption*{An example of applying the tug-the-hook symmetry at $N=0$. The \textcolor{bittersweet}{fat hook} is now of shape \textcolor{bittersweet}{$(2|2)$} and the symmetry may be thought as "tugging" the diagrams around the corner inside the fat hook. The diagrams are now parametrised by ordinary Frobenius variables:
}
$$
\hspace{10mm} R = (5, 3 \, | \, 4, 2) \hspace{15mm} \T_{-1}(R) = (4, 2 \, | \, 5, 3) \hspace{10mm} 
\T_{-2}(R) = (3, 1 \, | \, 6, 4)
$$
\end{figure}
Note that for the Alexander polynomial the tug-the-hook symmetry is "box preserving", i.e. it relates diagrams with equal $|R|$.\\\\
We would like to give various justifications for the claim and ways to think about this property. Important for us are the consequences of such a symmetry on the group-theoretical data of the HOMFLY invariant. \\\\
There are at least two ways of thinking about this property:
\begin{itemize}
\item It is a property of knot polynomials of \textbf{supergroup Chern-Simons theory}
\item It is a specific \textbf{incarnation of the eigenvalue conjecture}
\end{itemize}
\section{Supergroup knot invariants}
Knot invariants originating from supergroups have been quite extensively studied . As usual they can be approached in two ways: studying supegroup Chern-Simons theory \cite{Mikhaylov:2014aoa,Mikhaylov:2015qik} or quantum invariants given by $U_q(\sln(N|M))$ \cite{QS,Viro}.
\\\\
Supergroup Chern-Simons theory was in described in detail in \cite{Mikhaylov:2014aoa,Mikhaylov:2015qik}. The main idea is that $SU(N)$ Chern-Simons theory might be (as always with a lot of peculiarities) generalized to $SU(M+N|M)$ Chern-Simons theory. As discussed in \cite{Mikhaylov:2015qik} defining pure supergroup Chern-Simons theory is somewhat unclear, which does not stop us from making our conjecture. \\\\
As in usual Chern-Simons, the interesting observables are the Wilson loops:
\begin{equation}
\mathcal{W}_{R}(K)=\operatorname{Str}_{R} P \exp \left(-\oint_{K} \mathcal{A}\right)
\end{equation}
Putting all technicalities aside, we are interested in two main properties of these knot invariants:
\begin{enumerate}
\item The Wilson loop is labeled by a representation $R$ of the supergroup $SU(N|M)$.
\item The knot invariant for $SU(N|M)$ is equal (possibly up to factors) to that of CS theory with group $SU(|N-M|)$.
\end{enumerate}
The first claim is just a part of the definition. The second one can be justified in a number of ways, which are described in \cite{Mikhaylov:2015qik}. In short, in pure Chern-Simons it boils down to saying, that the Wilson loop expectation value is insensitive to the $U(1|1)$ factors. 
\\\\
The same is true from the quantum supergroup point of view. In fact in \cite{QS} it has been proven, that $U_q(\sln(N|M))$ knot polynomials are equal to those of $U_q(\sln(N-M))$
\\\\
Let us recall the very basics of representation theory for $\mathfrak{sl}(N|M)$ \cite{BERELE1987118,berele1983}. Representations of the superalgebra $\mathfrak{sl}(N|M)$ are labeled by partitions of the form:
\begin{equation}
\lambda_1 \geq \lambda_2 \geq \ldots \lambda_N \geq \lambda_{N+1} \geq \ldots
\end{equation}
where $\lambda_{i>N} \leq M$.  The partitions correspond to the highest weight representation as for the usual $\mathfrak{sl}(N)$. However, in the superalgebra case there is no restriction on the number components. The associated Young diagram should fit into a $(N|M)$ \emph{fat hook}:
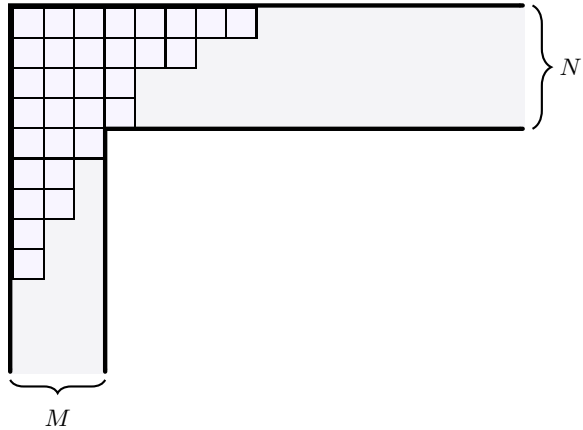
\begin{figure}[h]
    \centering
\begin{tikzpicture}
\node (1) at (0,0){
\ytableausetup{boxsize=1em}
\ydiagram[*(magnolia)]{8,6,4,4,3,2,2,1,1}};
\draw [line width=0.15em,black,round cap-round cap] 
([xshift=0.6em,yshift=-5cm]1.north west)--
([xshift=0.6em,yshift=-0.3em]1.north west) -- ([xshift=7cm,yshift=-0.3em]1.north west);
\draw [line width=0.15em,black,round cap-round cap] 
([xshift=3.245em+0.6em,yshift=-5cm]1.north west)--
([xshift=3.245em+0.6em,yshift=-4.55em]1.north west) -- ([xshift=7cm,yshift=-4.55em]1.north west);
\begin{scope}[on background layer]
\fill [anti-flashwhite,opacity=0.8] ([xshift=0.6em,yshift=-0.3em]1.north west) rectangle ([xshift=7cm,yshift=-4.55em]1.north west);
\fill [anti-flashwhite,opacity=0.8] ([xshift=0.6em,yshift=-0.3em]1.north west) rectangle ([xshift=3.245em+0.6em,yshift=-5cm]1.north west);
\end{scope}
\draw [thick,decorate,decoration={brace,amplitude=7pt,mirror},xshift=2em]
([xshift=7.1cm,yshift=-4.55em]1.north west) -- ([xshift=7.1cm,yshift=-0.3em]1.north west) node [black,midway,xshift=+0.5cm] 
{\footnotesize $N$};
\draw [thick,decorate,decoration={brace,amplitude=5pt,mirror},xshift=2em]
([xshift=0.6em,yshift=-5.08cm]1.north west) -- ([xshift=3.245em+0.6em,yshift=-5.08cm]1.north west) node [black,midway,yshift=-0.5cm] 
{\footnotesize $M$};
\end{tikzpicture}
\end{figure}
\\\\
It is an established fact, that for superalgebras Young diagrams  do not describe the representation uniquely. Particularly, a diagram with $N+\epsilon$ columns of $M$ rows, is equivalent to a diagram with $N$ columns of $M+\epsilon$ rows \cite{Bars:1982se}. 
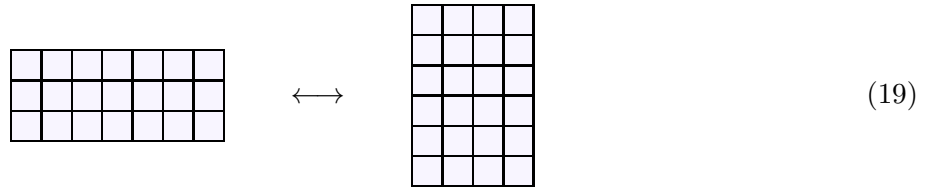
\begin{equation}\label{repequivalence}
\ytableausetup{boxsize=1em,aligntableaux = center}
\ydiagram[*(magnolia)]{7,7,7}
\qquad \longleftrightarrow \qquad
\ydiagram[*(magnolia)]{4,4,4,4,4,4}
\end{equation}
The described properties of supergroup knot invariants lead us to our conjecture. Mainly, we start by thinking about the normalized colored HOMFLY polynomials for $A=q^N$ defined as Wilson loops in representation of $\sln(N)$ given by a diagram $R$. Normalized HOMFLY polynomials for $R$ with $l(R)>N$  as $\sln(N)$ invariants lie in the \emph{non-physical region}. Meaning that formally such representations do not exist. These invariants, however, still make sense and are finite polynomials, because of the normalization factor in the definition \eqref{NormalizedHOMFLY}. They can be calculated in this region using for example, the $\mathcal{R}$-matrix approach, see section \ref{4}. 

On the other hand one can think of the Wilson loop at $A=q^N$ colored with the same diagram $R$ but in supergroup CS theory. Then $R$ is considered as a representation of $\sln(N+M|M)$ for some $M$ . The symmetry \eqref{sym} of the polynomial is then a remnant of the corresponding equivalence of representations \eqref{repequivalence}. Notice, that for fixed representation $R$ we obtain a variety of equalities suitably choosing $M$ and $N$. Surely we must admit, that these arguments are a vague abuse of some rigorous statements, which, however leads to an experimentally justified equality. Precise claims will be composed elsewhere. 

It is important to stress, that even though the symmetry relates HOMFLY polynomials in the non-physical region, it allows to conjecture results about the \textbf{general group theoretical structure of the invariant}.  Hence we would like to focus on the consequences of the statement.
\paragraph{Examples.}
First, let us give examples. We start with $N > 0$. These are some of the examples that we have checked for various knot like: torus - $[2,3]  \ , [3,4]$, non-torus - $4_1, 6_3$. For some classes of knots like torus and 3-strand HOMFLY polynomials in certain classes of representations can be calculated and the identity can be checked experimentally.
\begin{itemize}
\item The simplest non-trivial example is $R=[1,1,1]$.  We can only take $N=1,M=1,\epsilon=1$, meaning we are considering $\sln(2|1)$ representations. We have the following equality
\begin{align}\label{example1}
\Hn^{\K}_{[1,1,1]}(q,q^1)&=\Hn^{\K}_{[2,2]}(q,q^1) \qquad \ytableausetup{boxsize=0.5em,aligntableaux = center}\ydiagram{1,1,1} \longleftrightarrow \ydiagram{2,2} 
\end{align}
\item $R=[1^4]$. There are two options. Treating it as a representation of $\sln(3|1)$ with $\epsilon=1$ we get:
\begin{align}\label{example2}
\Hn^{\K}_{[1^4]}(q,q^2)&=\Hn^{\K}_{[2,2,2]}(q,q^2) \qquad \ytableausetup{boxsize=0.5em,aligntableaux = center}\ydiagram{1,1,1,1} \longleftrightarrow \ydiagram{2,2,2} 
\end{align}
While taking $\sln(2|1)$ with $\epsilon=2$:
\begin{align}\label{example3}
\Hn^{\K}_{[1^4]}(q,q^1)&=\Hn^{\K}_{[3,3]}(q,q^1) \qquad \ytableausetup{boxsize=0.5em,aligntableaux = center}\ydiagram{1,1,1,1} \longleftrightarrow \ydiagram{3,3} 
\end{align}
\item Finally we demonstrate a non-rectangular case. Take $R=[3,2]$, and consider it as $[2+1,2]$ for $\sln(2,1)$ and $\epsilon=1$. Then:
\begin{align}\label{example4}
\Hn^{\K}_{[3,2]}(q,q^1)&=\Hn^{\K}_{[2,1,1]}(q,q^1) \qquad \ytableausetup{boxsize=0.5em,aligntableaux = center}\ydiagram{3,2} \longleftrightarrow \ydiagram{2,1,1} 
\end{align}
\end{itemize}
\subsection{Perturbative analysis of the Alexander polynomial}
Now we would like to separately consider the limit $N=0$. From the point of view of $\sln(N)$ Chern-Simons theory it makes sense only as an "analytic continuation", whereas in the supergroup approach it is just an $\mathfrak{sl}(K|K)$ invariant (we ignore the comlications arising in representation theory of $\sln(K|K)$). The HOMFLY polynomial for $A=1$ is the so called Alexander polynomial \cite{Mironov:2016mxn}:
\begin{align}
\A_R^\K(q)=H_R^\K(q,A=1)
\end{align} 
It is named after the standart topological Alexander polynomial \cite{cromwell_2004}, which it is in the fundamental representation. 
The Alexander polynomial still has the tug-the-hook symmetry \eqref{AlexTTH}. It appears to be very constraining, when combined with the single-hook property:
\begin{equation}
\A_R^\K(q)=\A_R^\K(q^{|R|}), \text{for } R=[r,1^L]
\end{equation}
An extensive study of the single-hook property and it's relation to integrability was carried out in \cite{MIRONOV2018268,Mishnyakov:2019otq}. Arguments outlined in these papers explain how such "non-perturbative" properties affect the structure of the Vassilliev expansion (Kontsevich integral) \cite{Labastida:1997uw,chmutov_duzhin_mostovoy_2012}.
\\\\
Namely, one considers the expansion:
\begin{equation}
\mathcal{H}^{\K}_R(q=e^{\h},a=e^{N\h})=\sum_n \left( \sum_m v^\K_{nm} r^{R}_{nm} \right) \h^n
\end{equation}
where $v^\K_{nm}$ are the Vassiliev invariants, and $r^{R}_{nm}$ are the group factors. Any property  of $\mathcal{H}^\K_R(q,A)$ with respect to representations descends to group factors and becomes a condition on elements of $Z(U(\mathfrak{sl}_N))$. Therefore we get:
\begin{align}
r^R_{nm}\Big|_{N=0}  =r^{\T_\epsilon(R)}_{nm}\Big|_{N=0} &\text{ -  tug-the-hook symmetry,} \\
\left. r_{nm}^{[r,1^L]}=\left|R\right|r_{nm}^{[1]}\right|_{N=0} &\text{ - single-hook property.}
\end{align}
Being elements $Z(U(\mathfrak{sl}_N))$, they expand into a basis of shifted symmetric functions (eigenvalues of Casimir invariants):
\begin{align}
C_{r}(R)&=\sum_{i=1}\left(R_{i}-i+1 / 2\right)^{r}-(-i+1 / 2)^{r} \\ 
C_\Delta&=\prod_{i} C_{\Delta_i} , \ \Delta=[ \Delta_1,\Delta_2,\ldots ,\Delta_k]
\end{align}
 Hence, we can pose a general problem of finding polynomials in Casimir invariants, that satisfy the above relations:
\begin{align}
\sum_{|\Delta|=n}\alpha_\Delta C_\Delta(R)&=0, \  \text{for }  R=[r,1^L] \\
\label{tugthehook}\sum_{|\Delta|=n}\alpha_\Delta C_\Delta(R)&=\sum_{|\Delta|=n}\alpha_\Delta C_\Delta(\T_\epsilon(R))
\end{align} 
This will determine the subspace of $Z(U(\mathfrak{sl}_N))$ where the group factors lie. We can hope, and it appears so, that these symmetries are constraining enough to \textbf{fix the general possible form of the group factors for $\mathbf{N=0}$} to a large extent. \\\\
The system \eqref{tugthehook} was studied in \cite{Mishnyakov:2020khb}, where a way to construct an explicit solution and a combinatorial enumeration of the basis in the space of solutions is provided.\\\\
Hence extending the symmetry to the HOMFLY polynomial opens a problem of solving similar equations to determine already the group factors of the invariant for general $N$. This problem will be extensively studied in a future paper. However we want to make clear the general idea, that we expect to extract from the symmetry. Mainly, we hope to \textbf{determine the group theoretical dependence of the HOMFLY invariant in the perturbative expansion for general $\mathbf{N}$}.
\\\\
Our attention to supergroup formulation of CS theory and knot invariants was also motivated by a series of papers, which discussed the fundamental Alexander polynomial as a supergroup quantum invariant form various approaches:
\begin{itemize}
\item In the braid group/$\mathcal{R}$-matrix formalism for $U_q(\sln(N|M))$ and $U_q(\sln(1|1))$ \cite{Rozansky:1992zt,Viro,Reshetikhin2012} in particular in \cite{Kauffman:1990ix}, where it was also noticed that the various formulas for $U_q(\sln(N|M))$, like the values of Casimir invariants, skein relation, etc. looked as ones for $U_q(\sln(N-M))$.
\item In the conformal block approach for $SL(1|1)$ WZW model \cite{Rozansky:1992rx}, and the Drinfield associator approach \cite{Lieberum2002TheDA}.
\item As a Kontsevich integral \cite{Geer2005TheKI} and via superalgebra weight systems \cite{Figueroa-OFarrill:1996zvg}
\item As a version of categorification for $U_q(\sln(1|1))$ in \cite{Robert:2019yie}.
\end{itemize}
\section{Eigenvalue conjecture}\label{4}
We have argued that the tug-the-hook symmetry can be justified by considering the HOMFLY polynomial as a supergroup invariant. Now we would like to focus on the interplay of the symmetry with the Reshetikhin-Turaev approach to quantum knot invariants \cite{Reshetikhin:1990pr}. \\\\
We work with the HOMFLY polynomial as a $U_q(\sln(N))$ invariant. An interesting obsevation is that the tug-the-hook symmetry can be shown to hold, assuming the validity of the so called \textbf{eigenvalue conjecture}\cite{Itoyama:2012re}. This conjecture has not been proven yet, but there is a solid collection of arguments for its validity. Thus the proof of the tug-the-symmetry adds a new example to this collection . We will give a bried summary of the statement of the conjecture.\\\\
Begin with the $\mathcal{R}$-matrix expression for the normalized HOMFLY polynomial:
\begin{equation}\label{exp}
\Hn_R^{\K}=\sum_{Q \in R^{\otimes n}} \sigma_Q(\R) \dfrac{\qdim(Q)}{\qdim(R)} 
\end{equation}
Here $\sigma_Q$ are traces of products of $\mathcal{R}$ matrices in the multiplicity space of representation $Q$, $\qdim(Q)$ - is the quantum dimension of this representation \cite{Liu:2007kv}, finally the sum is taken over all $Q$ appearing in the decomposition of the tensor product $R^{\otimes n}$ into irreducible representations. The quantum dimension is given by a well known $q$-deformation of the hook-formula: 
\begin{equation}
\label{qdim}
    \qdim(Q) =  \prod_{(i, j) \in Q} \frac{[N - i + j]}{[h_{ij}]}.
\end{equation}
$$ h_{ij} := Q_i - i + Q^{\prime}_j - j + 1, \hspace{20mm} [n] := \frac{q^{n} - q^{-n}}{q - q^{-1}}.$$
The symmetry \eqref{sym} interplays with each part of the expression:  the irreducible components $Q$, quantum dimensions and $\sigma_Q(\mathcal{R})$. In order to establish the role our statement plays in such approach we will treat each constituent of \eqref{exp} separately. 
\\
In fact the quantum dimension is exactly the value of the HOMFLY polynomial of the unknot:
\begin{equation}
H^\bigcirc_R(q,A)= \qdim(R)
\end{equation}
so \eqref{exp} is a formula for \eqref{NormalizedHOMFLY}.
\paragraph{Action on irreducible respresentations.} The foremost question to ask is whether the decomposition into representtions $Q$ has some nice property with respect to applying our symmetry to $R$? Let us recall the simple case, corresponding to the Alexander polymomial $N=0$, which was described in \cite{Mishnyakov:2020khb
}. \\\\
The sum over $Q$ in \eqref{exp} is determined by the Littlewood-Richardson rule. However we also have the ratio of quantum dimension, which can evaluate to zero in some cases. The denominator of \eqref{qdim} never vanishes, but what about the numerator? For $N=0$ the formula becomes:
\begin{align}\label{qdimN0}
\qdim(R)\Big|_{N=0}=\prod_{i,j}\dfrac{[j-i]}{[h_{ij}]}
\end{align}
Clearly the quantum dimensions have zeros on the diagonal boxes of the Young diagram. Hence the only way for the ration to be non-vanishing, one has to have the same number of the diagonal boxes, i.e. the same number of hooks, in $Q$ and in $R$. Therefore in formula \eqref{exp} the same actually goes over the set:
\begin{align}
Q \in R^{\otimes n} : Q \vdash n|R| , \  h(Q) = h(R) , \text{where } h(R) =  \text{\{number of hooks in the diagram\}}
\end{align}
The symmetry amounts to transforming $R \rightarrow \T_{\epsilon}(R)$, while the subrepresentations $Q$ will just transform accordingly:
\begin{align}
\T_{n\epsilon}(Q) \in (\T_{\epsilon}R)^{\otimes n} : Q \vdash n|R| , \  h(Q) = h(R)
\end{align}
Therefore we observe a nice action of the symmetry on one part of formula \eqref{exp}. \\\\
The $N \neq 0 $ case is slightly more complicated, since now the symmetry changes the number of boxes. In the hook formula for the quantum dimension the denominator evaluates to zero in boxes that lie on a diagonal, shifted downwards by $N$ rows. An example for $N=2$:
\begin{center}
\begin{tikzpicture}[remember picture,>=triangle 45]
\node (A) at(-1,0){
\ytableausetup{boxsize=2em}
\begin{ytableau}[*(magnolia)]
*(magnolia)& & & &  \\
& & & & \\
0& & & & \\
&0 & & \\
& &0 & \\
& & 
\end{ytableau}};
\node [above=0.3cm of A] (B) {An example of $Q=[5,5,5,4,4,3]$};
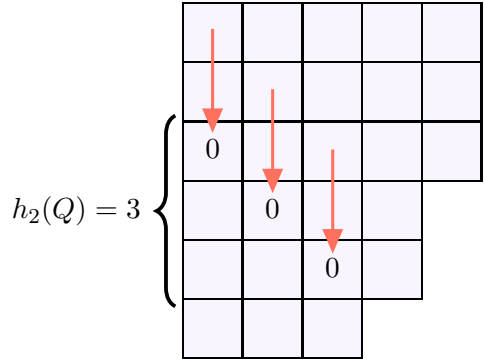
\node (u1) at (-0.93,0.55) {};
\node (u2) at (-0.93,-1.1) {};
\node (u3) at (-1.72, 1.35) {};
\node (u4) at (-1.72,-0.3) {};
\node (u5) at (-2.51, 2.15) {};
\node (u6) at (-2.51, 0.5) {};
\node (u7) at (-3, 1) {};
\node (u8) at (-3, -1.8) {};
\draw[->,line width=0.4mm,bittersweet] (u1) to (u2);
\draw[->,line width=0.4mm,bittersweet] (u3) to (u4);
\draw[->,line width=0.4mm,bittersweet] (u5) to (u6);
\draw[decorate,decoration={brace,mirror,amplitude=7pt},line width=0.5mm] (u7) -- (u8) node [midway, left=0.3cm of u8]{$h_2(Q)=3$};
\end{tikzpicture}
\end{center}

Denote by $h_N(R)$ the number of boxes on the shifted diagonal. Referring to the discussion in the previous sections, the number of these shifted hooks is exactly the $M$ in the rank of the supergoup $\sln(N,M)$. 

For the representation  $Q$ to give a non-vanishing contribution to \eqref{exp} it has to have $h_N(Q)= h_N(R)$ 
Then for the HOMFLY polynomial at $A=q^N$ we have the following representiations contributing to \eqref{exp}:
\begin{equation}
Q \in R^{\otimes n}: Q\vdash n|R|, \ h_N(Q)= h_N(R)
\end{equation}
After applying the symmetry the represetations transform in a simple way, similar to the $N=0$ case:
\begin{align}
\T^N_{n\epsilon}(Q) \in (\T^N_{\epsilon}R)^{\otimes n} : Q \vdash n|R| , \  h_N(Q)= h_N(R)
\end{align}
Now let us deal with the quantum dimensions. 
\paragraph{Invariance of the quantum dimension}
It appears that the quantum dimensions are invariant under the action of the symmetry on the representation up to a sign: 
\begin{align}\label{qdiminv}
\dfrac{\qdim(\T^N_\epsilon(R))}{\qdim(R)}=(-1)^{\epsilon h_N(R)}
\end{align}
\begin{Remark}
We remind that it is crucial to deal with the ratio $\frac{\qdim(\T^N_\epsilon(R))}{\qdim(R)}$ as a whole. This is because formally both are zero in $\sln(N)$ for such "long" diagrams. The ratio, however are finite.
\end{Remark} 
As an illustration consider one of the examples above \eqref{example4}:
\begin{align}
\dfrac{\qdim([3,2])}{\qdim([2,1,1])}=\dfrac{[1]\cdot[2]\cdot[3]\cdot[0]\cdot[1]}{[4]\cdot[3]\cdot[1]\cdot[2]\cdot[1]}\cdot\dfrac{[4]\cdot[1]\cdot[2]\cdot[1]}{[1]\cdot[2]\cdot[0]\cdot[-1]}=-1.
\end{align}
We observe that it is indeed invariant. Since this paper is intended to be qualitative, we won't give a proof here. It is a simple combinatorial calculation, the $N=0$ case is provided in the appendix to \cite{Mishnyakov:2020khb}.
\paragraph{$\mathcal{R}$-matrices.}
According to the general theory $\sigma_Q(R)$ are characters of the braid group representation given by $\R$-matrices. If we denote by $\pi(\beta^\K)$ the representation fo the braid group correspoding to the knot, then:
\begin{align}
\sigma_Q(\R)=\mathrm{\phantom{}_q tr}(\pi(\beta^\K))
\end{align}
The representation is given by matrices $\R_i$. The eigenvalue conjecture states that \textbf{the quantum $\mathcal{R}_i$-matrices are fully determined be the eigenvalues of the universal $\mathcal{R}$ matrix}.\\\\ This means that it is enough to prove the invariance of the full set of eigenvalues under the action of the symmetry. The eigenvalues of the universal $\mathcal{R}$-matrix are determined by the tensor square of two $U_q(\mathfrak{sl}(N))$ representations:
\begin{equation}
R \otimes R = \bigoplus_{|W|=2|R|}  W 
\end{equation}
The eigenvalues $\lambda_W$ for the each irriducible component are given by:
\begin{align}
\lambda_W=q^{\kappa(W)-4\kappa(R)-N|R|}, \\
\kappa(R)=2\sum_{i,j \in R}(j-i) =C_2(R)
\end{align}
When we apply the symmetry, the new set of eigenvalues will be determined by the decomposition of the transoformed representations:
\begin{equation}
\T^N_\epsilon (R) \otimes \T^N_\epsilon (R) = \bigoplus_{|W|=2|R|} \T^N_{2\epsilon} (W) 
\end{equation}
The fact that the irreducible subrepresentations transform in a nice way under the action of the tug-the-hook symmetry leads to the conservation of the whole set of eigenvalues, due to the following equality:
\begin{equation}\label{preserv}
\begin{split}
\kappa(W)-4 \kappa(R)-N|R|&=\kappa\left(\T^N_{2\epsilon} (W)\right)-4\kappa\left(\T^N_\epsilon (R)\right)-N\left|\T^N_\epsilon (R) \right| \\
\text{for }h_N(R)&=h_N(W)
\end{split}
\end{equation} 
This means, that the set of eigenvalues is preserved up to a sign, we can schematically wright:
\begin{equation}
\{\lambda\}_R=\{\lambda\}_{\T^N_\epsilon(R)}
\end{equation}
This means that the traces  are also preserved. A more carefull consideration shows that the sign exactly cancels the sign appearing from the quantum dimension:
\begin{equation}
\sigma_{\T^N_\epsilon(W)}(\mathcal{R})=(-1)^{\epsilon h_N(R)}\sigma_{W}(\mathcal{R})
\end{equation}
Hence we see, that due to the properties of the components in the expression \eqref{exp} with respect to the tug-the-hook symmetry we can prove the invariance of the HOMFLY polynomial.

\section{Conclusion/Discussion}
In the paper we presented a new symmetry of the colored $\mathfrak{sl}(N)$ HOMFLY polynomial, based on ideas about supergroup Chern-Simons theory. It extends the tug-the-hook symmetry of the colored Alexander polynomial. We aimed at demonstrating two points. 

First, is that for the Alexander polynomial the perturbative treatment of such type of symmetries leads to strong restrictions on the group theoretical components. This means that an analogous treatment is possible in the general case. The efficiency of the symmetries in determining the group factors of the Alexander group factors gives a hope of finding some general expressions for the HOMFLY group factors and/or the dimensions of the corresponding spaces. Speaking in QFT terms we at least partially solve the field theory data of the CS Wilson loops, leaving only the purely topological part undetermined.

Secondly, we showed, that the extended tug-the-hook symmetry interplays in a fashionable manner with the eigenvalue conjecture. It acts in a natural way on each constituent of the quantum trace formula and can be basically proved provided the eigenvalue conjecture is valid. We see this, as another strong argument in favor of the conjecture, since the symmetry appears absolutely independently in the supergroup CS context and can be proven for quantum supergroups.

This paper outlines the main ideas and statement. Most of those require rigorous proofs which we intend to provide in longer version of this paper in the future.
There is a large variety of further questions that this symmetry poses for future studies, some of them are:
\begin{itemize}
\item Finding general solutions to the symmetry property in terms of Casimir invariants and defining correctly the corresponding ideal in $Z\left(U(\mathfrak{sl}(N)\right)$
\item Describing the integrability properties these group factors might have as in the Alexander case.
\item Possible applications in various questions in the theory of quantum knot invariants: homological invariants, knot-quiver correspondence, differential expansion, WZW conformal blocks approach to Chern-Simons theory.
\end{itemize}
\section{Acknowledgments}
This work was funded by the Russian Science Foundation (Grant No.16-12-10344).
\bibliographystyle{unsrturl} 
\bibliography{alexnewsym}

\end{document}